\documentclass[twocolumn,10pt,aps,amsmath,amssymb,pra]{revtex4-1}

\usepackage{amsmath}
\usepackage{amssymb}
\usepackage{bm}
\usepackage[usenames,dvipsnames]{color} 
\usepackage{cancel}
\usepackage{dcolumn}
\usepackage{epsf}
\usepackage{epsfig}
\usepackage{graphicx}
\usepackage{mathtools}
\usepackage{multirow}
\usepackage[final]{pdfpages}
\usepackage{setspace}
\usepackage{ulem} 
\usepackage{wrapfig}

\newcommand\Id{\leavevmode\hbox{\small1\normalsize\kern-.33em1}}
\newcommand{\half}{\frac{1}{2}}

\newcommand{\ave}[1]{\left\langle #1\right\rangle}

\newcommand{\ham}{{\mathcal{H}}}
\newcommand{\sx}{\sigma_x}
\newcommand{\sy}{\sigma_y}
\newcommand{\sz}{\sigma_z}

\definecolor{DarkBlue}{rgb}{0,0,.4}
 
\newcommand{\blist}[1]{
 \begin{list}{#1}
  { \setlength{\itemsep}{3pt}
     \setlength{\parsep}{2pt}
     \setlength{\topsep}{3pt}
     \setlength{\partopsep}{0pt}
     \setlength{\leftmargin}{1em}
     \setlength{\labelwidth}{1em}
     \setlength{\labelsep}{0.5em} } }
\newcommand{\elist}{
  \end{list}  }

\DeclareMathSymbol{\vartheta}{\mathalpha}{letters}{"12}
\DeclareMathSymbol{\theta}{\mathalpha}{letters}{"23}
\DeclareMathSymbol{\phi}{\mathalpha}{letters}{"27}
\DeclareMathSymbol{\varphi}{\mathalpha}{letters}{"1E}

\definecolor{LinkColor}{rgb}{0,0,.5}
\usepackage[colorlinks=true,linktoc=BrickRed,linkcolor=BrickRed,citecolor=LinkColor,urlcolor=LinkColor]{hyperref}

\newcommand{\SW}{\,\mathbb{SW}}
\renewcommand{\emph}{\textit}

\graphicspath{{../../../figures/},{/Users/paola/Documents/Work/figures/}}
\begin{document}
\title {Continuous dynamical decoupling magnetometry}
\author{Masashi Hirose}
\email{These authors contributed equally to this work.}
\author{Clarice D.~Aiello}
\email{These authors contributed equally to this work.}
\author{Paola Cappellaro}
\email{pcappell@mit.edu}
\affiliation{Nuclear Science and Engineering Department, Massachusetts Institute of  Technology, Cambridge, MA 02139, USA}
\begin{abstract}
Solid-state qubits hold the promise to achieve unmatched combination of sensitivity and spatial resolution. To achieve their potential, the qubits need however to be shielded from the deleterious effects of the environment. While dynamical decoupling techniques can improve the coherence time, they impose a compromise between sensitivity and bandwidth, since to higher decoupling power  correspond higher frequencies of the field to be measured. Moreover, the performance of pulse  sequences is ultimately limited by control bounds  and errors. 
Here we analyze a  versatile alternative  based on continuous driving. We find that continuous dynamical decoupling schemes can be used for AC magnetometry, providing similar frequency constraints on the AC field and improved sensitivity for some noise regimes. In addition, the flexibility of phase and amplitude modulation could yield  superior robustness to driving errors and a better adaptability to external experimental scenarios. 
\end{abstract}
\maketitle

Solid-state qubits have emerged as promising quantum sensors, as they can be fabricated in small volumes and  brought close to the field to be detected. Notably, Nitrogen-Vacancy (NV) centers in nano-crystals of diamond~\cite{Jelezko02} have been applied for high sensitivity detection of magnetic~\cite{Taylor08,Maze08,Balasubramanian08} and electric fields~\cite{Dolde11} and could be used either as nano-scale scanning tips~\cite{Maletinsky12} or even in-vivo due their small dimensions and low cytotoxicity~\cite{McGuinness11}.  Unfortunately, solid-state qubits are also sensitive probes of their environment~\cite{Bar-Gill12,Bylander11} and this leads to rapid signal decay, which limits the sensor interrogation time and thus its sensitivity. 
Dynamical decoupling (DD) methods~\cite{Carr54,Viola99b,Uhrig07,Khodjasteh07,Biercuk11} have been adopted to prolong the coherence time of the sensor qubits~\cite{Taylor08,deLange11,Bar-Gill12,Pham12}. 
Although DD techniques prevent measuring constant, DC fields, they provide superior sensitivity to oscillating AC fields, as they can increase the sensor coherence time by orders of magnitude.  The sensitivity is maximized by carefully matching the  decoupling period to the AC field; conversely, one can study the response of a decoupling scheme to fields of various frequencies, thus mapping out their bandwidth. Still, the refocusing power of pulsed DD techniques is ultimately limited by pulse errors and bounds in the driving power. 
Here we investigate an alternative strategy, based on continuous dynamical decoupling (CoDD), that has the potential to overcome these limitations.

We consider the problem of measuring  a small external field, coupled to the sensor by a Hamiltonian: $\ham_{b}=\gamma b(t) S_{z}$,
where $S_{z}$ is the spin operator of the quantum sensor.  For example, $b(t)$ can be an external magnetic field and $\gamma$  the spin's gyromagnetic ratio. The figure of merit for a quantum sensor is the smallest field $\delta b_{min}$ that can be read out during a total time $\mathbf{t}$, that is, the sensitivity $\eta=\delta b_{min}\sqrt{\mathbf{t}}$. We use this metric to compare pulsed and continuous DD schemes and show how CoDD can offer an advantage for some noise regimes.

\begin{figure}[t]
\centering
\includegraphics[width=0.45\textwidth]{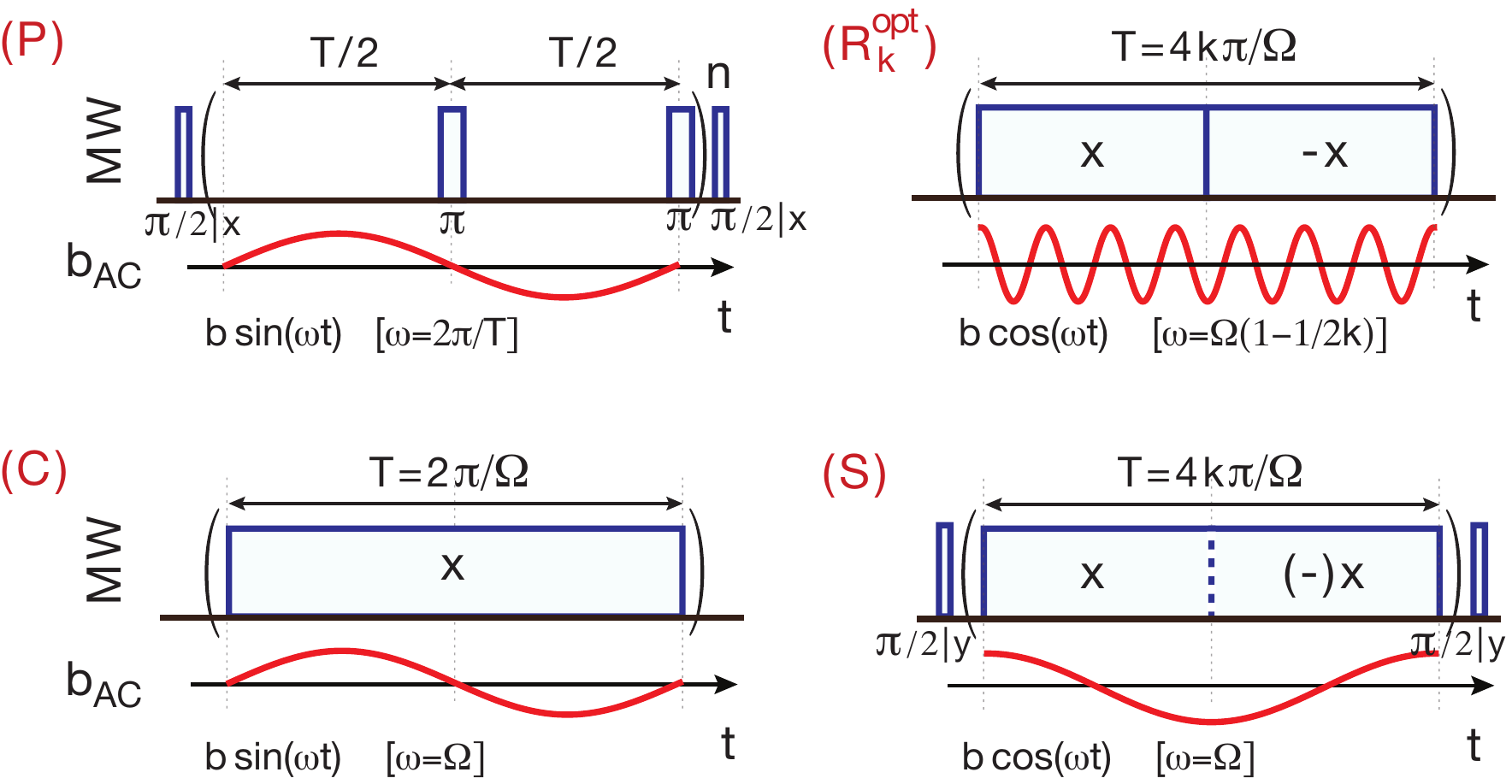}
\caption{Pulse sequences for four AC magnetometry schemes:  PDD (P), constant driving (S), RE with optimal frequency (R$_k^{\text{opt}}$) and spin-locking (S). Blue boxes represent microwave driving, with phase  (x and y) as indicated.}
\label{fig:Sequence}\vspace{-12pt}
\end{figure}

The principle of DD schemes rests on the spin echo sequence,  which refocuses unwanted phase accumulation due to a slow bath by reversing the system evolution with control pulses.  More complex DD sequences can in principle extend the coherence time indefinitely, by increasing the number of pulses. 
In practice, however,  a large number of imperfect, finite-width pulses provokes the accumulation of error and degrades DD performance~\cite{Khodjasteh07,Khodjasteh05,Wang12}. 
CoDD has been first introduced in the context of NMR  to mitigate  pulse errors~\cite{Burum81,Boutis03}  and it has then lead to many schemes, such as composite pulses~\cite{Shaka83b,Levitt86}, dynamically corrected gates~\cite{Khodjasteh09} and optimized modulations~\cite{Jones12}. In general, phase and amplitude modulation of the continuous driving allows great flexibility and CoDD can achieve high decoupling power.
Here we consider only two schemes, constant continuous driving (C) and  Rotary Echo (RE)~\cite{Solomon57,Aiello12,Laraoui11}, as their periodicity allows an easier use for AC magnetometry (see Fig.~\ref{fig:Sequence}); we will compare these schemes to the simplest pulsed DD scheme, period dynamical decoupling (PDD).

As an example, we compute the signal and sensitivity of AC magnetometry under RE, but similar derivations apply for the other schemes. The RE  sequence consists of a continuous on-resonance driving field of constant amplitude $\Omega$ and phase inverted at periodic intervals (see Fig.~\ref{fig:Sequence}). RE is parametrized by the angle $\theta=\Omega T/2$, where $T$ is the sequence period. While RE is usually employed to refocus errors in the driving field, for $\theta=2\pi k$ the sequence also refocus dephasing noise, with performance depending on both $k$ and the Rabi frequency. 
We consider the evolution of a sensor qubit under a sequence of $2\pi k$-RE and in the presence of an external AC magnetic field of frequency $\omega$ whose magnitude $b$ is to be sensed:
\begin{equation}
\mathcal{H}(t) = \Omega \mathbb{SW}(t)S_x + \gamma b\cos(\omega t + \phi)S_z,
\end{equation}
where $\mathbb{SW}(t)$ is the square wave of period $T = {4\pi k}/{\Omega}$. In the toggling frame of the driving field, the Hamiltonian becomes
\begin{equation}
\widetilde{\mathcal{H}}(t)\!=\!\frac{\gamma b\cos(\omega t+\phi)}{2}[ \cos(\Omega t)S_z-\mathbb{SW}(t)\sin(\Omega t)S_y ].
\end{equation}
We consider only the cases where $\phi=0$ and $\omega T=2m\pi$, with $m$ an odd integer, since as we show below this yields good sensitivities. 
 Under this assumption $\tilde{\mathcal{H}}(t)$ is periodic and for small fields $b$ the evolution operator can be well approximated from a first order average Hamiltonian over the  period $T$, $\overline{\mathcal{H}} \approx \frac{1}{T}\int_{0}^{T}\tilde{\mathcal{H}}(t)dt=\gamma\overline b\,S_y$.

If $m = 1$, we define $\omega_{low} = \frac{\Omega}{2 k}$, which, for a fixed $\Omega$, is easily adjustable by changing the echo angle $2\pi k$. 
Setting instead $m = (2k-1)$, we define $\omega_{opt} = \frac{\Omega (2k-1)}{2k}$, which yields $\overline b=4bk/[\pi(4k-1)]$ and attains the best sensitivity of the method. 
The sensitivity,  obtained as $\eta(t) = \displaystyle\lim_{b \rightarrow 0}\textstyle\frac{\Delta\mathcal{S}}{|\frac{\partial \mathcal{S}}{\partial b}|}\sqrt{t}$, where $\mathcal{S}$ is the signal and $\Delta\mathcal{S}$ its shot-noise limited uncertainty, depends on $\overline b$, that is, on the averaging of the AC field over the sequence  period due to the DD modulation. 
We compare the performance of both $2\pi k$-RE schemes to   PDD  (optimum $\omega = {2\pi}/{t}$, $\phi = {\pi}/{2}$) and a constant modulation with $\omega=\Omega$ (see Fig.~\ref{fig:Sequence}). 
We obtain for the   schemes considered:
\begin{equation}
\renewcommand{\arraystretch}{2}
\begin{array}{lclc}
\eta^{opt}_{R_k}=\eta\frac{4k-1}{2k}     &\quad (\theequation.a) &\qquad \eta_P=\eta&\quad (\theequation.b)\\
\eta^{low}_{R_k}=\eta\frac{4k^2-1}{2k} &\quad (\theequation.c) &\qquad \eta_C=\frac4\pi\eta&\quad (\theequation.d),
\end{array}\nonumber \label{eq:sensitivity}
\addtocounter{equation}{1}\end{equation}
where $\eta=\frac{\pi }{2\gamma C \sqrt{t}}$, with  $C$ a parameter capturing inefficiencies in the sensor readout~\cite{Taylor08}.  Here $R_k$ labels a $2k\pi$-RE scheme, $P$ the PDD scheme and $C$ the constant modulation (see Figure~\ref{fig:Sequence}).  A fourth operating scheme can be obtained by a ``spin-locking'' sequence~\cite{Redfield55}, where the spin is first rotated to the transverse plane before applying a driving field in the same direction; choosing $\phi=0$ and $\omega=\Omega$ yields the same sensitivity as for the constant modulation, $\eta_S\!=\!\eta_C$, even when the driving phase is inverted periodically. 
We note that if the phase $\phi$ of the AC field is not optimized, the sensitivities are reduced by a factor $\Phi(\phi)$, with $\Phi_P=\Phi_C=\csc(\phi)$ and $\Phi_{R_k}=\sec\phi$. If in addition the  phase of the AC field cannot be fixed, $\Phi(\phi)=\sqrt{2}$ when considering the average signal over many realizations.

These ideal sensitivities are degraded in the presence of noise and whenever the frequency of the AC field is not matched to the DD period. In the following we analyze these two contributions, showing that they lead to a sensitivity  $\eta\to\eta\mathcal D(t)/W(\omega)$, where $\mathcal D(t)$ describes the decay under DD sequences and $W(\omega)$ is the reduction in the accumulated phase  when the field frequency $\omega$ is suboptimal.

\begin{figure}[t]
\centering
\includegraphics[width=0.22\textwidth]{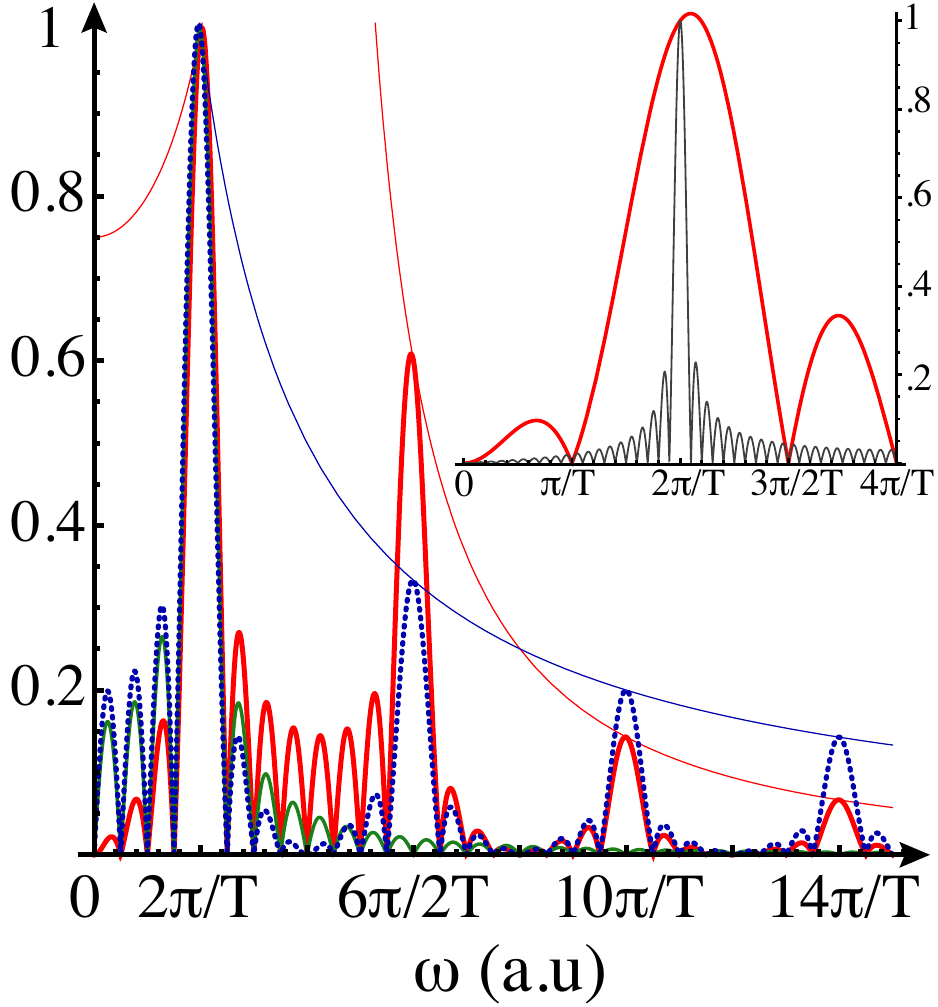}\quad\includegraphics[width=0.22\textwidth]{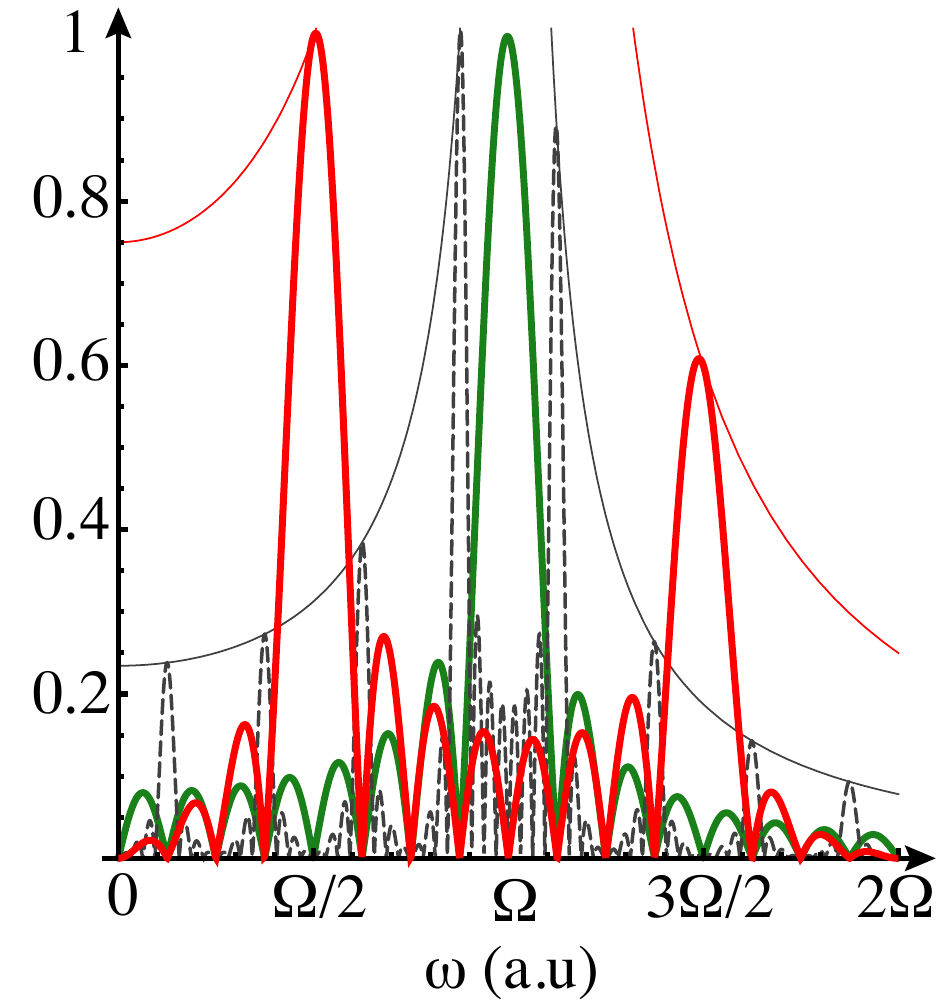}
\caption{Bandwidth for AC magnetometry. We plot the  weight functions $W(\omega)$ that scale the phase acquired during DD magnetometry for AC fields of frequency $\omega$.
Left: we plot $W(\omega)$ and the envelop of its passband decay  for PDD (blue dotted), RE ($k=1$, red, thick) and constant driving (green, thin  line) for $n=2$ cycles, expressing the frequency in terms of the sequence period. In the inset: we compare the main peak for $n=1$ (red, thick) and $n=10$ (gray) for RE ($k=1$) showing the reduction in bandwidth.
Right: we compare $W(\omega)$ for continuous driving (green) and for RE with $k=1$ (red, thick) and $k=4$ (gray, dashed), plotting as a function of $\omega$ in units of the Rabi frequency $\Omega$.}\vspace{-12pt}
\label{fig:bandwidth}
\end{figure}

Optimal sensitivities are obtained by carefully matching the period of the DD schemes to  the oscillating field.  In practice, however,  when field frequencies are either unknown or known to a finite precision, it is of relevance to determine the bandwidth of the scheme and the deviation from optimum sensitivities. 
We estimate the bandwidth by calculating the phase accumulated by the sensor over the total interrogation time $t = nT$, $\overline B t= \int_{0}^{t}b(t) f(t)\mathrm{d}t$, and examining the frequency dependence of its absolute value.  For PDD, the filter function is $f_{P}(t) =\mathbb{SW}_{P}(t)$,  the square wave with the period of the modulation. For continuous driving schemes such as RE and Rabi, $f(t)$ is the strength of the toggling frame Hamiltonian. In particular, $f_{R_k}(t) = \mathbb{SW}(t)\sin(\Omega t)$ 
yielding the weight function $W_{R_k}(\omega)=|\overline B_{R_k}(\omega)|/|\overline B_{R_k}(\omega^{opt})|$:
\begin{equation}
W_{R_k}(\omega)\!=\!\frac{(4 k-1)/n}{\left|(4 k)^2-\left(\frac{T \omega }{\pi }\right)^2\right|}\left|\sin (n T \omega ) \tan \left(\frac{T \omega }{4}\right)\right| .
\end{equation}
$W_{R_k}$ has  peaks (\textit{pass-bands}) at $\omega = {2\pi(2(k+p)-1)}/{T}$, where $p$ is an integer satisfying $p\geq1-k$. The lowest pass-band occurs for $p = 1-k$, corresponding to $\omega_{low} = {\Omega}/{2k}$. The strongest peak is for $p=0$ at $\omega_{opt}$. Subsequent periodic peaks are attenuated from the symmetry point $\omega = \Omega$ as $\sim \frac{\Omega^2}{|\omega^2 - \Omega^2|}$.
The FWHM of the optimum peak in $W_{R_k}(\omega)$ decays as $\approx \frac{7.58}{2nT}$, where $7.58 \approx$ FWHM of the squared sinc function, a result common to the other DD schemes.

A similar calculation for the accumulated phase during a PDD sequence indicates the existence of peaks at $\omega = {m\pi}/{T}$, with $m$ odd,  whose intensity decays as $1/m$. This slower  decay than for the RE pass-bands could be beneficial if the goal is to detect fields of unknown frequencies. 
On the other hand,  AC magnetometry under continuous driving or spin locking could be used for frequency-selective detection because $W_{C}(\omega)$ has a unique peak at $\omega = {2\pi}/{T}$ with  FWHM  on the same order of that for RE. A comparison of the different weight functions  is depicted in Fig.~\ref{fig:bandwidth}. We note that while $W(\omega)$ describes the poor performance of DD schemes at detecting AC fields with unmatched frequencies, this property could in turn be used for frequency-selective measurements and even spectroscopy, by scanning the sequence period. While constant driving provides the best selectivity (canceling out higher octaves), RE provides more flexibility by changing both the period time and the angle $2\pi k$, which allows more uniform noise cancellation.

\begin{figure}[t]
\centering
\includegraphics[width=0.42\textwidth]{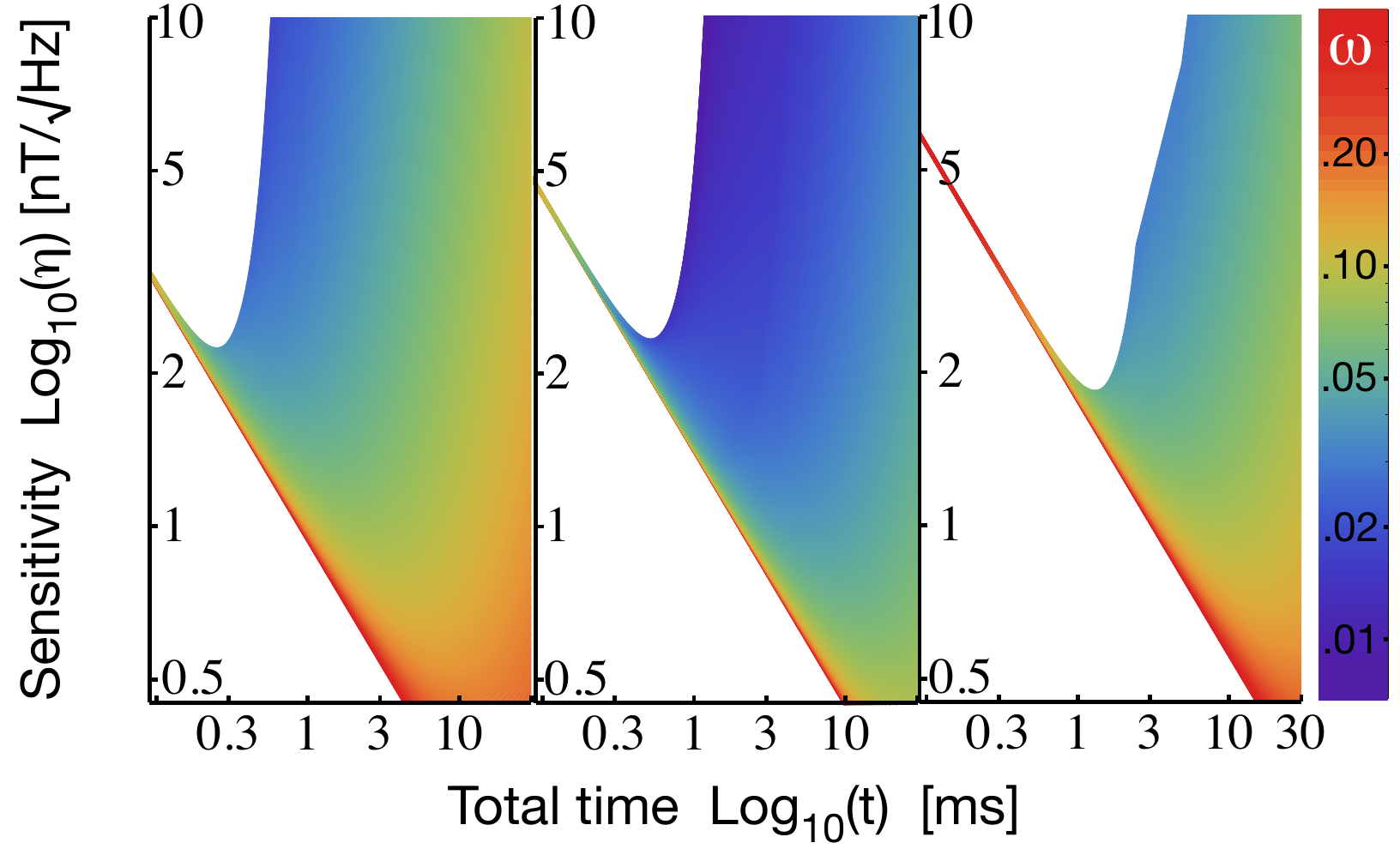}
\caption{Sensitivity for AC magnetometry. We compare the magnetic field sensitivity of a single NV center for PDD (left) and RE ($k\!=\!1$ center; $k\!=\!4$ right). We assumed  $T_2\!=\!500\mu$s under OU noise (comparable to a $^{13}$C bath), yielding a decay $\propto\!e^{-T^3/(n^2 T_2^3)}$, and a single readout with $C=0.03$. A larger number of refocusing cycles (with shorter periods) achieves better sensitivity but can only detect higher frequencies, as shown by the color of the curves (right bar, MHz).}\vspace{-12pt}
\label{fig:SensitivityACtime}
\end{figure}
 The refocusing power of RE can surpass that of pulsed schemes.  Consider for example a noise with long correlation time $\tau_c$: In this limit, the signals decays as 
$\ave{\mathcal{S}_{R_k}(t)}=e^{-(\Gamma_{2R}t)^3/n^2}$, with $\Gamma^3_{2R}=\frac{3\sigma^2}{8 k^2 \pi ^2 \tau_c}$.
Using a  similar derivation~\cite{Kubo62,Cappellaro06}, the  decay under  a  PDD sequence is instead $\ave{\mathcal{S}_P(t)}=e^{-(\Gamma_{2P}t)^3/n^2}$, with $\Gamma^3_{2P}=\frac{2\sigma^2}{3 \tau_c}$. 
The sensitivities in Eq.~(\ref{eq:sensitivity}) are further limited by the signal decay $\mathcal D(t)$ under the DD sequences. The achievable sensitivity is then a compromise between the refocusing power of the sequence used and the frequency that it allows detecting (Fig.~\ref{fig:SensitivityACtime}).
While the decay for pulsed DD has been widely studied, evolution under continuous DD is more complex~\cite{Dobrovitski09}.
We can estimate the RE decay to first leading order using a cumulant expansion~\cite{Kubo62,Cappellaro06}. We assume a stochastic Hamiltonian, $\ham(t)=\Omega \SW_k(t)\sx+\delta(t)\sz$, where $\delta(t)$ is an Ornstein-Uhlenbek noise with zero mean and autocorrelation function $G(\tau)=\sigma^2 e^{-\frac{\tau}{\tau_c}}$, with  $\sigma$ the dispersion and  $\tau_c$ the correlation time. 
The signal decay can be calculated from the average of the superoperator $\ave{\mathbb S(t)}=\ave{\mathcal T e^{-i\int_0^t\widehat Hdt'}}$, where we indicate by a hat the superoperators $\widehat A=A\otimes\Id-\Id\otimes A$ and  $\mathcal T$ is the time ordering operator.
In turns, this can be approximated by the cumulants, $\ave{\mathbb S(t)}\approx \exp{[-(K_1+K_2+\dots)t]}$, with the first cumulant $K_1=0$ and the second given by 
\[K_2=\frac1{2t}\int_0^tdt_1\,\int_0^{t_1}dt_2\,\langle{\widehat H(t_1)\widehat H(t_2)}\rangle_c,\] 
 where the cumulant average is
 \[\langle{\widehat H(t_1)\widehat H(t_2)}\rangle_c=\mathcal T\langle\widehat H(t_1)\widehat H(t_2)\rangle-\langle{\widehat H(t_1)}\rangle\langle{\widehat H(t_2)}\rangle.\]
In the toggling frame of the driving field, the stochastic Hamiltonian is $\tilde \ham(t)=\delta(t)N(t)\equiv\delta(t)\left[\cos(\Omega t) \sz+ \SW \sin(\Omega t)\sy\right]$. 
Then the second cumulant for $n$ cycles is 
$K_2\!=\!n\triangle+\square\sum_{j=1}^n(n-j) G_j$ \cite{Cappellaro06}, with $G_j=e^{-\frac{4k\pi j}{\Omega\tau_c}}$ and
 \[\triangle=\int_0^{4k\pi/\Omega}dt_1\,\int_0^{t_1}dt_2  {\widehat N}(t_1){\widehat N}(t_2) G(t_1-t_2),\]
 \[\square=\int_0^{4k\pi/\Omega}dt_1\,\int_0^{2k\pi/\Omega}dt_2  {\widehat N}(t_1){\widehat N}(t_2) G(t_1-t_2).\]
 The cumulant  can be written as 
 \begin{equation}
 K_2=\frac{\alpha+\beta}2\hat S_z^2+\frac{\alpha-\beta}2\hat S_y^2+\frac{\sqrt{\gamma^2-\beta^2}}2(\hat S_y\hat S_z+\hat S_z\hat S_y)
 \label{eq:K2}
 \end{equation} 
 (see appendix for explicit expressions), yielding the signal  
\[\ave{\mathcal{S}_{R_k}} =\half[1+\mathcal D_R]=\half\left[1+e^{-\alpha} (  \cosh(\gamma)+\frac{\beta}{\gamma}  \sinh(\gamma))\right].\]
Numerical simulations match well with these approximate analytical results. 

\begin{figure}
\centering
\includegraphics[width=0.4\textwidth]{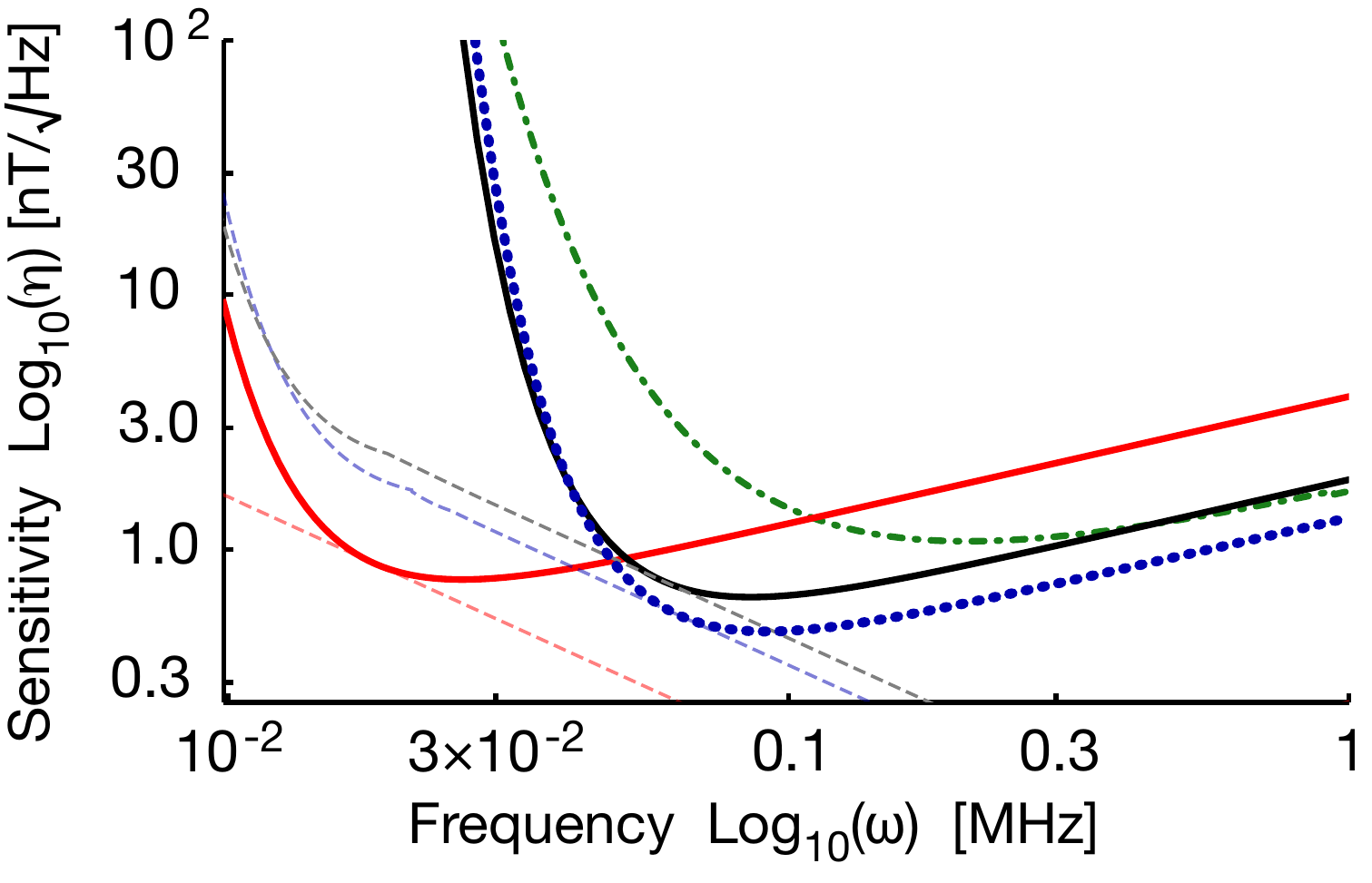}
\caption{Sensitivity for AC magnetometry. We compare the achievable sensitivity for constant driving (green, dash-dotted), PDD of $n=50$ echoes (blue, dotted) and RE ($2\pi$-RE, red, achieving the same sensitivity of PDD at a lower frequency and $8\pi$-RE, black, achieving better sensitivity than PDD at the same frequency). We assumed  $T_2=500\mu$s under OU noise, yielding a super-exponential decay $\propto e^{-T^3/(n^2 T_2^3)}$, and a single readout with $C=0.03$. The decay of the constant (Rabi) driving was calculated following Ref.~\cite{Dobrovitski09} for long  $\tau_c$. The dashed, thin lines correspond to the ideal limit with no driving or pulse errors.}
\label{fig:SensitivityACFreq}
\end{figure}

The longer coherence time under the RE sequence can be exploited either to reach a better sensitivity for a given frequency or to measure lower frequency fields at a given sensitivity, as shown in figure~\ref{fig:SensitivityACFreq}. The achievable improvement depends on the effective coherence time ratio, $\tau=T_{2R}/T_{2E}$, obtained from the two schemes. Because of the improved refocusing of RE with respect to PDD, the sensitivity can be improved for some noise regimes. In addition, RE-AC magnetometry provides the flexibility of using larger angles (larger $k$) to allow for longer interrogation times (Figure~\ref{fig:SensitivityACtime}) at lower frequencies, which could be beneficial in practical cases in combination with repeated readout schemes~\cite{Neumann10b,Aiello12}.

We remark that besides the decay functions obtained above in the presence of dephasing noise, other sources of decay can arise from imperfect pulses or fluctuations in the driving power. To this effect,  RE provides a good protection against slow fluctuation in the driving power~\cite{Solomon57, Aiello12} and it is thus expected to achieve much better overall sensitivities than a continuous driving.
In conclusion, we analyzed a novel scheme for AC magnetometry based on continuous dynamical decoupling and compared its performance to pulsed DD schemes. While we focused on the simplest DD sequences, we note that more complex driving, such as composite pulses~\cite{Levitt86,Aiello12}, could achieve even better refocusing of driving field instability and inhomogeneity while still providing comparable sensitivity.
We further analyzed the response of AC magnetometry to fields of unknown frequencies, finding that some CoDD schemes (such as continuous driving or spin locking with alternating phases) are advantageous for spectroscopy.  The  sensitivity is ultimately limited not only by the theoretically achievable coherence time, but also by pulse errors or fluctuations in the driving field. 
While a full comparison of the limits due to imperfection in the control fields is beyond the scope of this work, the flexibility of CoDD schemes in modulating both phase and amplitude of the driving field can provide practical advantages, yielding a better compromise between the DD refocusing power and the frequencies of the field to be measured.  
\\
\textbf{Acknowledgments} This work was supported in part by  the
ARO through  grant No. 
W911NF-11-1-0400 and by DARPA. 
C.~D.~A. 
acknowledges support from the Schlumberger Foundation.
\appendix
\begin{widetext}
\section{Cumulant}
We can calculate the time (ensemble) average of a time-ordered exponential operator by means of a cumulant expansion
The first cumulant is zero since we assume a zero-average  as zero.  The second cumulant for the RE sequence is given by Eq.~\ref{eq:K2} with 
\begin{equation}\begin{array}{ll}
\alpha=&\frac{\sigma ^2 T^2 \tau _c e^{-\frac{n T}{\tau _c}}}
{\left(e^{\frac{T}{2 \tau _c}}+1\right)^2 \left(16 \pi ^2 k^2 \tau _c^2+T^2\right)^2}
\left[2 n \left(e^{\frac{T}{2 \tau _c}}+1\right){}^2 e^{\frac{n T}{\tau _c}} \left(16 \pi ^2 k^2 T \tau _c^2+64 \pi ^2 k^2 \tau _c^3 \tanh \left(\frac{T}{4 \tau _c}\right)+T^3\right)\right.\\
&\left.-8 \tau _c e^{\frac{(n+1) T}{2 \tau _c}} \left(\left(T^2-16 \pi ^2 k^2 \tau _c^2\right)+\left(16 \pi ^2 k^2 \tau _c^2+T^2\right) \cosh \left(\frac{T}{2 \tau _c}\right)\right) \sinh \left(\frac{n T}{2 \tau _c}\right)\right]\end{array}\end{equation}
\begin{equation}\begin{array}{ll}
\beta=&-\frac{2 \sigma ^2 T^2 \tau _c^2 e^{-\frac{n T}{\tau _c}}}
{\left(e^{\frac{T}{2 \tau _c}}+1\right)^2 \left(16 \pi ^2 k^2 \tau _c^2+T^2\right)^2}\times
\\&\left[16 \pi ^2 k^2 \tau _c^2 \left(e^{\frac{T}{2 \tau _c}}-1\right) \left(e^{\frac{n T}{\tau _c}} \left((4 n-1) e^{\frac{T}{2 \tau _c}}+4 n+1\right)+e^{\frac{T}{2 \tau _c}}-1\right)+T^2 \left(e^{\frac{T}{2 \tau _c}}+1\right){}^2 \left(e^{\frac{n T}{\tau _c}}-1\right)\right]\end{array}\end{equation}
\begin{equation}\begin{array}{ll}
\gamma=&-\frac{2 \sigma ^2 T^2 \tau _c^2 e^{-\frac{n T}{\tau _c}} }
{\left(e^{\frac{T}{2 \tau _c}}+1\right)^2 \left(16 \pi ^2 k^2 \tau _c^2+T^2\right)^2}
\left[64 \pi ^2 k^2 T^2 \tau _c^2 \left(e^{\frac{T}{2 \tau _c}}+1\right)^4 \left(e^{\frac{n T}{\tau _c}}-1\right)^2 \tanh\left(\frac{T}{4 \tau _c}\right)^2+\right.\\
&\left.\left(16 \pi ^2 k^2 \tau _c^2 \left(e^{\frac{T}{2 \tau _c}}-1\right) \left(e^{\frac{n T}{\tau _c}} \left((4 n-1) e^{\frac{T}{2 \tau _c}}+4 n+1\right)+e^{\frac{T}{2 \tau _c}}-1\right)  + T^2 \left(e^{\frac{T}{2 \tau _c}}+1\right)^2 \left(e^{\frac{n T}{\tau _c}}-1\right)\right)^2\right]^{1/2}\end{array}\end{equation}
\end{widetext}

\bibliography{../../Biblio}

\end{document}